\documentstyle[aps,prb,amsfonts,amssymb,graphicx,epsfig,preprint]{revtex}
\draft
\title{How well do Car-Parrinello simulations reproduce the Born-Oppenheimer 
surface? Theory and examples.}

\author{P. Tangney\footnote{Current address : 
International School for Advanced Studies,
         via Beirut 2-4, 34013 Trieste, Italy.\\
E-mail address: tangney@sissa.it} and S. Scandolo\footnote{Current address: Dept. of Chemistry, 
Princeton University,
Princeton, New Jersey 08540}}

\address{Princeton Materials Institute,
Dept. of Chemistry and Dept. of Geosciences,\\ 
Princeton University,
Princeton, New Jersey 08540 \\
International School for Advanced Studies, 
         via Beirut 2-4, 34013 Trieste, Italy.\\
Istituto Nazionale per la Fisica della Materia, via Beirut 2-4, 34013 Trieste, Italy}
\begin{document}
\maketitle
\begin{abstract}
We derive an analytic expression for the average difference between the 
forces on the ions in a Car-Parrinello simulation
and the forces obtained at 
the same ionic positions when the electrons are at their ground state. 
We show that for common values of the fictitious electron mass,
a systematic bias may affect the Car-Parrinello forces in systems 
where the electron-ion coupling is large.
We show that in the limit where the electronic orbitals are rigidly dragged 
by the ions the difference between the two dynamics amounts to a 
rescaling of the ionic masses, thereby leaving the thermodynamics intact.
We study the examples of crystalline magnesium oxide and crystalline and 
molten silicon. We find that for crystalline silicon the errors are very 
small. For crystalline MgO the errors are very large but the 
dynamics can be quite well corrected within the rigid-ion model. 
We conclude that it is important to control the effect of the electron mass
parameter on the quantities extracted from Car-Parrinello simulations. 
\end{abstract}
\newpage
\section{Introduction}
Since its introduction in 1985 the Car-Parrinello (CP)
method \cite{cp} has increasingly been used to study an 
ever wider range of problems in the dynamics and thermodynamics 
of solids and liquids under various conditions 
and in studying the dynamics of chemical reactions.
CP is an efficient method to solve the Kohn-Sham equations 
of Density Functional Theory\cite{ks} ``on the fly'', as the electronic 
ground state evolves due to changing ionic positions. 
It is based on the introduction of an additional inertia 
associated with the electronic orbitals, which are evolved 
as classical degrees of freedom along the ionic Molecular Dynamics 
(MD) trajectory. Its popularity stems from its efficiency 
relative to full Born-Oppenheimer (BO) dynamical methods, 
where the electronic orbitals are forced to be in the ground 
state for each ionic configuration, and from the observation
that apart from small fluctuations which average out 
on a femtosecond timescale the physical 
quantities which are extracted from it are indistinguishable 
from BO dynamics\cite{buda}. 

It is known that the introduction into the electronic system 
of a fictitious inertia introduces differences into the dynamics 
relative to the BO dynamics. However these differences have never
been fully quantified or analysed in the context of the 
appropriate theoretical framework.
In this paper we carefully examine the relationship 
between CP and BO dynamics and show how the difference between them
scales with the value of the fictitious
mass parameter, $\mu$.
We derive analytic expressions for such differences in the limit 
of the fictitious kinetic energy associated with electronic orbitals 
being a minimum.
As has been proposed previously\cite{blochl2}, 
we show how these errors may be corrected
in the ideal case where the electron-ion coupling 
can be modeled as a rigid dragging of localized atomic orbitals and
show its application to specific examples. 
We suggest that quantities extracted from CP simulations should always be
checked against their possible dependence on the value of $\mu$.

The central issue is the following: the classical motion of the 
electronic orbitals in CP can be thought of as being made up of 
two components. One component consists of fast oscillations with period
equal to or faster than $\tau_e \sim 2\pi\sqrt{\mu/E_g}$, $E_g$ being the 
lowest electronic excitation 
energy and $\mu$ the fictitious electronic mass\cite{buda}. 
The second component is the unavoidable but ``adiabatic'' response of 
the electronic orbitals to the ionic dynamics, whose shortest characteristic 
time we denote by $\tau_i$ ($\tau_e \ll \tau_i$). It is generally thought 
that keeping the time scales (or frequency spectra) of the two components 
well separated (i.e. by reducing $\mu$) ensures a correct adiabatic decoupling 
and guarantees that the CP dynamics is a faithful representation of the 
BO dynamics. We show here that the adiabatic decoupling is a necessary 
but not sufficient condition for the accuracy of CP dynamics. 
In particular, the fictitious inertia also causes 
the slow component of the electronic dynamics to exchange momentum and energy with 
the ions, yielding a departure of the CP forces on the 
ions from the BO ones for large values of $\mu$. 

We begin by outlining the relevant theory and deriving an expression
for the difference between CP forces and BO forces in the limit of
minimum fictitious electronic kinetic energy. We then consider the
simplified model of rigid ions and derive the error in the forces for this
system. 
We show how the thermodynamics and the dynamics can be corrected
for systems in which the rigid-ion model provides a good description
of the electronic structure. 
We illustrate these theoretical considerations
by studying the examples of Magnesium Oxide and Silicon. 

\section{Theory}
The Car-Parrinello method makes use of the following classical lagrangean :
\begin{equation}
L_{CP} = \sum_{i}\mu_{i} \langle\dot{\psi}_{i}|\dot{\psi}_{i}\rangle
+ \frac{1}{2}\sum_{I}M_{I}\dot{\textbf{R}}_{I}^{2} 
- E[\{\psi_{i}\},\{\textbf{R}_{I}\}] \label{eqn:lagrange1}
\end{equation}
to generate trajectories for the ionic and electronic 
degrees of freedom via the coupled set of equations of motion
\begin{equation}
\textrm{M}_{I}\ddot{R}_{I}^{\alpha} = 
-\frac{ \partial E [\{\psi_{i}\},\{\textbf{R}_{I}\}]}
{\partial R_{I}^{\alpha}}= F_{CP_{I}}^{\alpha}\label{eqn:ions1}
\end{equation}

\begin{equation}
\mu_{i}|\ddot{\psi}_{i}\rangle = 
-\frac{\delta E[\{\psi_{i}\},\{\textbf{R}_{I}\}]}{\delta 
\langle \psi_{i}|} \label{eqn:electrons1}
\end{equation}
where $M_{I}$ and ${\textbf{R}_{I}}$ are the mass and position 
respectively of atom $I$, $|\psi_{i}\rangle$ are
the Kohn-Sham orbitals which are allowed to evolve as 
classical degrees of freedom with inertial parameters  $\mu_{i}$, 
and $E[\{\psi_{i}\},\{\textbf{R}_{I}\}]$
is the Kohn-Sham energy functional evaluated for the set of 
ionic positions $\{\textbf{R}_{I}\}$ and the set of orbitals $\{\psi_{i}\}$.
The functional derivative of the Kohn-Sham energy in (\ref{eqn:electrons1})
is implicitly restricted to variations of $\{\psi_{i}\}$ that preserve
orthonormality.

We wish to compare the dynamics of ions evolved with this method 
with the true BO dynamics.
For this purpose, we decompose the CP orbitals as
\begin{equation}
|\psi_{i}\rangle = |\psi_{i}^{(0)}\rangle + |\delta\psi_{i}\rangle \label{eqn:decomp}
\end{equation}
where $|\psi_{i}^{(0)}\rangle$ are the ground state (BO) orbitals 
which are uniquely defined for given ionic coordinates as those that
minimize $E [\{\psi_{i}\},\{\textbf{R}_{I}\}]$ . This allows us to consider 
separately the evolution of the instantaneous electronic ground state and 
the deviations of the CP orbitals from that ground state. 

A preliminary interesting observation follows from such a decomposition:
strictly speaking, the CP equations do not reduce to the BO equations in the
limit of vanishing $\delta\psi_{i}$, for any finite value of $\mu$. In fact,
because of the dependence of the ionic coordinates on time, the BO 
orbitals have a non vanishing ``acceleration'' given by
\begin{equation}
|\ddot{\psi}_{i}^{(0)}\rangle 
= \sum_{I}\ddot{R}_{I}^{\alpha}
\frac{\partial|\psi_{i}^{(0)}\rangle}{\partial R_{I}^{\alpha}}
+ \sum_{I,J} \dot{R}_{I}^{\alpha}\dot{R}_{J}^{\beta}\frac{\partial^{2}|
\psi_{i}^{(0)}\rangle}{\partial R_{J}^{\beta}\partial R_{I}^{\alpha}}
\label{eqn:kick}
\end{equation}
where we have used the fact that
\begin{equation}
|\dot{\psi}_{i}^{(0)}\rangle=\sum_{I}\dot{R}_{I}^{\alpha}
\frac{\partial|\psi_{i}^{(0)}\rangle}{\partial R_{I}^{\alpha}}
  \label{eqn:orbvel}
\end{equation}
It now becomes clear that Eqn. (\ref{eqn:electrons1}) is not compatible
with a vanishing departure of the CP orbitals from the BO orbitals,
since the right-hand side of (\ref{eqn:electrons1}) would vanish 
if $\{\psi_{i}\} = \{\psi_{i}^{(0)}\}$, while the left-hand side would not,
by virtue of (\ref{eqn:kick}). 
So the CP orbitals cannot take their ground state values unless
$\mu$ vanishes too. As a consequence of this, the ionic dynamics 
is affected by a bias proportional to $\mu$ and, as we will see, 
to the strength of the electron-ion interaction .

We now wish to explore the consequences that such a departure 
from the ground state has
on the instantaneous CP forces $F_{CP}$. We thus calculate 
how CP forces deviate from the BO forces $F_{BO}$ at 
a given point in phase space along the CP trajectory.
We may write, for the $\alpha$-th cartesian component of the force on atom $I$:
\begin{eqnarray}
- F_{CP_{I}}^{\alpha} =
\frac{\partial E[\{\textbf{R}_{I}\};\{\psi_{i}\}]}{\partial R_{I}^{\alpha}} & = & 
\frac{d E[\{\textbf{R}_{I}\};\{\psi_{i}\}]}{d R_{I}^{\alpha}} \nonumber \\ & - & 
\sum_{i}\bigg (\frac{\delta E[\{\textbf{R}_{I}\};\{\psi_{i}\}]}{\delta |\psi_{i}\rangle} 
\frac{\partial |\psi_{i}^{(0)}\rangle}{\partial R_{I}^{\alpha}}
+\frac{\partial \langle \psi_{i}^{(0)}|}{\partial R_{I}^{\alpha}}
\frac{\delta E[\{\textbf{R}_{I}\};\{\psi_{i}\}]}{\delta \langle \psi_{i}|} \bigg) \label{eqn:fcp1}
\end{eqnarray}
Substitution of equation ~\ref{eqn:electrons1} yields
\begin{equation}
- F_{CP_{I}}^{\alpha} =
\frac{d E[\{\textbf{R}_{I}\};\{\psi_{i}\}]}{d R_{I}^{\alpha}} - 
\sum_{i}\mu_{i}\bigg (\langle\ddot{\psi}_{i}|         
\frac{\partial |\psi_{i}^{(0)}\rangle}{\partial R_{I}^{\alpha}}
+\frac{\partial \langle \psi_{i}^{(0)}|}{\partial R_{I}^{\alpha}}
|\ddot{\psi}_{i}\rangle \bigg) \label{eqn:fcp2}
\end{equation}
Using the expansion
\begin{eqnarray}
\frac{d E[\{\textbf{R}_{I}\};\{\psi_{i}\}]}{d R_{I}^{\alpha}} & = & 
\frac{d}{d R_{I}^{\alpha}}\bigg\{E[\{\textbf{R}_{I}\};\{\psi_{i}\}]
\Bigg |_{\{\psi_{i}^{(0)}\}}  \nonumber \\
& + &\sum_{i}\bigg (\frac{\delta E[\{\textbf{R}_{I}\};\{\psi_{i}\}]}
{\delta |\psi_{i}\rangle}\Bigg |_{\{\psi_{i}^{(0)}\}}|\delta \psi_{i}\rangle
+\langle \delta \psi_{i}|\frac{\delta E[\{\textbf{R}_{I}\};\{\psi_{i}\}]}
{\delta \langle \psi_{i}|}\Bigg |_{\{\psi_{i}^{(0)}\}} \bigg) + order(\delta \psi_{i}^{2}) 
\bigg \} \nonumber \\
& = & - F_{BO_{I}}^{\alpha} + 0 + order(\delta \psi_{i}^{2})
\end{eqnarray}
we can write the error in the CP force as 
\begin{equation}
\Delta F_{I}^{\alpha}=F_{CP_{I}}^{\alpha} - F_{BO_{I}}^{\alpha} 
=  \sum_{i}\mu_{i}\bigg (\langle\ddot{\psi}_{i}|
\frac{\partial |\psi_{i}^{(0)}\rangle}{\partial R_{I}^{\alpha}}
+\frac{\partial \langle \psi_{i}^{(0)}|}{\partial R_{I}^{\alpha}}
|\ddot{\psi}_{i}\rangle \bigg) + order(\delta \psi_{i}^{2}) \label{eqn:error}
\end{equation}
Having established the connection, to first order in $\delta \psi_{i}$, between
the CP and the BO forces, we assume adiabatic decoupling and look 
for contributions to this difference
that do not vanish when averaged over time scales longer than the 
fictitious dynamics of the electrons ($\tau_e$) but shorter than the 
time scales of the ionic dynamics ($\tau_i$). 
Only these contributions are expected to contribute
significantly to the ionic dynamics\cite{buda}. 
To this end we rewrite Eqn. (\ref{eqn:electrons1}), 
using (\ref{eqn:decomp}), as
\begin{eqnarray}
|\ddot{\psi}_{i}\rangle & = & |\delta\ddot{\psi}_{i}\rangle + \sum_{I}\ddot{R}_{I}^{\alpha}\frac{\partial|\psi_{i}^{(0)}\rangle}{\partial R_{I}^{\alpha}}
+ \sum_{I,J} \dot{R}_{I}^{\alpha}\dot{R}_{J}^{\beta}\frac{\partial^{2}|\psi_{i}^{(0)}\rangle}{\partial R_{J}^{\beta}\partial R_{I}^{\alpha}} \nonumber \\
& = &-\frac{1}{\mu_{i}}\frac{\delta E[\{\psi_{i}\},\{\textbf{R}_{I}\}]}{\delta \langle \psi_{i}|} 
= -\frac{1}{\mu_{i}} \bigg( \frac{\delta^{2} E[\{\textbf{R}_{I}\},\{\psi_{i}\}]}{\delta | \psi_{i} \rangle
 \delta \langle \psi_{i}|}\Bigg |_{\{\psi_{i}^{(0)}\}} |\delta\psi_{i}\rangle 
+ order(\delta \psi_{i}^{2})\bigg)  \label{eqn:electrons3}
\end{eqnarray}
and we show that when eq. (\ref{eqn:electrons3}) is averaged 
over a time scale shorter than $\tau_i$ but longer 
than $\tau_e$ , then $\delta \psi_{i}$ vanishes. 
In order to prove it, we re-express equation ~\ref{eqn:electrons3} as 
\begin{eqnarray}
\langle\varphi_{j}|\delta\ddot{\psi}_{i}\rangle + 
\langle\varphi_{j}|\chi^{(i)}\rangle
\simeq -\frac{{k}_{j}^{(i)}}{\mu_{i}}
\langle\varphi_{j}|\delta\psi_{i}\rangle \label{eqn:electrons4}
\end{eqnarray}
where
\begin{eqnarray}
|\chi^{(i)}(\{\textrm{R}_{I}(t),\dot{\textrm{R}}_{I}(t),\ddot{\textrm{R}}_{I}(t)\}) \rangle = 
\sum_{I}\ddot{R}_{I}^{\alpha}\frac{\partial|\psi_{i}^{(0)}\rangle}{\partial R_{I}^{\alpha}}
+ \sum_{I,J} \dot{R}_{I}^{\alpha}\dot{R}_{J}^{\beta}
\frac{\partial^{2}|\psi_{i}^{(0)}\rangle}{\partial R_{J}^{\beta}\partial R_{I}^{\alpha}} 
\label{eqn:chi} \\
\hat{K}^{(i)}(\{\textrm{R}_{I}(t)\}) = \frac{\delta^{2} 
E[\{\textbf{R}_{I}\},\{\psi_{i}\}]}{\delta | \psi_{i} \rangle
 \delta \langle \psi_{i}|}\Bigg |_{\{\psi_{i}^{(0)}\}}
\end{eqnarray}
and $|\varphi_{j}^{(i)}\rangle$ is an eigenvector of $\hat{K}^{(i)}$ 
with eigenvalue $k_{j}^{(i)}$.
$|\chi^{(i)}\rangle$ and $\hat{K}^{(i)}$ , by definition, vary on the timescale of ionic motion. 
We consider timescales much smaller than $\tau_i  = 2\pi/\omega_i$ 
such that $|\chi^{(i)}\rangle$,$\hat{K}^{(i)}$ and
$|\varphi_{j}\rangle$ are approximately constant.  
In this case a solution of ~\ref{eqn:electrons4} 
is of the form
\begin{eqnarray}
\langle\varphi_{j}|\delta\psi_{i}\rangle = Be^{\iota\omega_{j}^{(i)} t}
- \frac{\langle\varphi_{j}|\chi^{(i)}\rangle}{(\omega^{(i)}_{j})^{2}}
\label{eqn:electrons5}
\end{eqnarray}
where $\omega_{j}^{(i)}= \sqrt{k_{j}^{(i)}/\mu_{i}}$ 
and $B$ is a complex constant.
If we assume that $\omega_{j}^{(i)}$ is very large relative to $\omega_{i}$ 
then the average value of $\langle\varphi_{j}|\delta\psi_{i}(t)\rangle$ 
over timescales much greater than $2\pi/\omega_{j}^{(i)}$ is
\begin{eqnarray}
\overline{\langle \varphi_{j}(\tau)| \delta\psi_{i}\rangle} & = & 
\frac{1}{\Delta\tau}\int_{\tau - \frac{\Delta \tau}{2}}^{\tau 
+ \frac{\Delta \tau}{2}}\langle \varphi_{j}(\tau)| \delta\psi_{i}(t)\rangle dt \nonumber \\
& \approx & - \frac{\langle\varphi_{j}(\tau)|\chi^{(i)}(\tau)\rangle}{(\omega_{j}^{(i)})^{2}}
\;\;\;\;\; {\textrm for} \;\;  \frac{2\pi}{\omega_{j}^{(i)}} 
\ll \Delta\tau \ll \frac{2\pi}{\omega_{\scriptstyle i}} \\
\Rightarrow \overline{\langle \varphi_{j}(\tau)|\delta\ddot{\psi}_{i}\rangle} 
& \approx & 0 \label{eqn:average}
\end{eqnarray}
Since 
\begin{eqnarray}
|\delta\ddot{\psi}_{i}\rangle = \sum_{j} \langle\varphi_{j}|\delta\ddot{\psi}_{i}\rangle
|\varphi_{j}\rangle
\end{eqnarray}
this means that for timescales which are intermediate between typical timescales of
ionic and CP orbital motion $\overline{|\ddot{\psi}_{i}\rangle} \approx |0\rangle$.

To summarise : If we consider the dynamics 
of the electronic orbitals to consist of an adiabatic 
response of the electronic orbitals to the ionic dynamics 
and an independent fast oscillating part then, under the 
assumption that the timescales of the fast component
are much shorter than the shortest time period in the ionic system, 
i.e. assuming adiabatic decoupling,
the average error in the Car-Parrinello forces is given by 
(using equations ~\ref{eqn:error},~\ref{eqn:electrons3} and ~\ref{eqn:average})

\begin{eqnarray}
\Delta F_{I}^{\alpha} = 2 \sum_{i}\mu_{i}\Re\bigg \{\sum_{J}\ddot{R}_{J}^{\beta}\frac{\partial \langle \psi_{i}^{(0)}|}{\partial R_{I}^{\alpha}}
\frac{\partial|\psi_{i}^{(0)}\rangle}{\partial R_{J}^{\beta}}
+ \sum_{J,K} \dot{R}_{J}^{\beta}\dot{R}_{K}^{\gamma}\frac{\partial \langle \psi_{i}^{(0)}|}{\partial R_{I}^{\alpha}}
\frac{\partial^{2}|\psi_{i}^{(0)}\rangle}{\partial R_{K}^{\gamma}\partial R_{J}^{\beta}}\bigg\}
+ order(\delta \psi_{i}^{2}) \label{eqn:error2}
\end{eqnarray}

\section{The Rigid Ion Approximation.}
In order to gain insight into the scale of this problem with the CP forces we consider the simple example of rigid ions.
We assume that each electron is localised around an ion and that there is no distortion of a particular 
ion's charge distribution as it moves in the field of the other ions. We can refer each wavefunction $\psi_{i}$ to a particular ion as follows
\begin{equation}
\psi_{i}(\textbf{r}) = \phi_{I\eta}(\textbf{r} - \textbf{R}_{I})
\end{equation}
Where the electronic states are labelled by an ion index, $I$,
  and the index $\eta$ labelling
the electronic state of the ion .
The rigidity of the ionic charge distribution means that
\begin{eqnarray}
\frac{\partial \phi_{I\eta}(\textbf{r} - \textbf{R}_{I})}
{\partial \textbf{R}_{I} } = -\frac{\partial \phi_{I\eta}(\textbf{r} - \textbf{R}_{I})}
{\partial \textbf{r}}
\qquad\textrm{and}\qquad
\frac{\partial \phi_{I\eta}(\textbf{r} - \textbf{R}_{I})}
{\partial \textbf{R}_{J} } = 0 \;\;\;\;\;\;\; \forall J \neq I
\label{eqn:rigid}
\end{eqnarray}
Equation ~\ref{eqn:error2} becomes
\begin{eqnarray}
\Delta F_{I}^{\alpha} & = & 2 \sum_{\eta}\mu_{\eta}\Re\bigg\{ \ddot{R}_{I}^{\beta}\int \frac{\partial \phi_{I\eta}^{*}(\textbf{r} - \textbf{R}_{I})}
{\partial r^{\alpha }}\frac{\partial \phi_{I\eta}(\textbf{r} - \textbf{R}_{I})}{\partial r^{\beta}}d\textbf{r} \nonumber \\
& + & \dot{R}_{I}^{\beta}\dot{R}_{I}^{\gamma}\int \frac{\partial \phi_{I\eta}^{*}(\textbf{r} - \textbf{R}_{I})}
{\partial r^{\alpha }} \frac{\partial^{2} \phi_{I\eta}(\textbf{r} - \textbf{R}_{I})}{\partial r^{\gamma}\partial r^{\beta}}d\textbf{r}\bigg \} \label{eqn:error3}
\end{eqnarray}
The second term in equation ~\ref{eqn:error3} vanishes due to symmetry, 
at least assuming an atomic charge density with spherical symmetry.
The first term may be written in terms of $E_{k}^{I\eta}$ the quantum 
electronic kinetic energy of an electron in state $\eta$ of atom $I$ as
\begin{eqnarray}
2 \sum_{\eta}\mu_{\eta}\Re\bigg\{ \ddot{R}_{I}^{\beta}\int \frac{\partial \phi_{I\eta}^{*}(\textbf{r} - \textbf{R}_{I})}
{\partial r^{\alpha }}\frac{\partial \phi_{I\eta}(\textbf{r} - \textbf{R}_{I})}{\partial r^{\beta}}d\textbf{r}\bigg\} = - \frac{2m_{e}}{3 \hbar^{2}} \ddot{R}_{I}^{\alpha}
\sum_{\eta} \mu_{\eta}E_{k}^{I\eta } \label{eqn:A}
\end{eqnarray}
where $m_{e}$ is the (real) mass of an electron.
Since the ions are rigid the quantum kinetic energy associated with each one is a constant and equation ~\ref{eqn:error3} becomes
\begin{eqnarray}
\Delta F_{I}^{\alpha} & = & - \Delta M_{I}\ddot{R}_{I}^{\alpha} \label{eqn:error4}
\end{eqnarray}
with
\begin{equation}
\Delta M_{I} = \frac{2m_{e}}{3 \hbar^{2}} \sum_{\eta} \mu_{\eta}E_{k}^{I\eta}
\label{eqn:deltaM}
\end{equation}
In this case the ionic positions and velocities are updated during a Car-Parrinello simulation 
according to
\begin{equation}
(M_{I}+\Delta M_{I})\ddot{R}_{I}^{\alpha} = F_{BO_{I}}^{\alpha} \label{eqn:masscorr}
\end{equation}
In other words, for systems where the rigid ion approximation is valid, the CP 
approximation amounts simply to a rescaling of the ionic masses. 
Since the classical partition function depends
only on the interaction potential, the thermodynamics of the system 
as calculated with a CP dynamics is identical to the thermodynamics 
of the BO system. The definition of temperature will however be affected,
because if the actual ionic dynamics in CP is given by (\ref{eqn:masscorr}),
then the real temperature at which the system equilibrates, at least 
in the case of a microcanonical dynamics for the ions, is given by 
\begin{equation}
k_B T = \frac{1}{3N} \sum_{I,\alpha}(M_{I}+\Delta M_{I}) 
                                    \langle(v_I^{\alpha})^{2}\rangle
\label{eqn:realtemp}
\end{equation}
where $\langle\cdots\rangle$ signifies the average over time and 
$N$ is the number of atoms. 
This differs from the standard definition by the addition of a term
proportional to $\Delta M_{I}$. 
The additional term in eq. (\ref{eqn:realtemp}) can be readily traced to
the additional inertia caused by the rigid dragging of the electronic 
orbitals. In fact, using (\ref{eqn:orbvel}) and (\ref{eqn:rigid}), we can show 
that this term coincides, within the rigid ion model, with the 
fictitious electronic kinetic energy, when the contribution from the 
dynamics of the $\delta\psi_{i}$ is negligible i.e.
\begin{equation}
T_{el} 
 = 
\sum_{i}\mu_{i} \langle\dot{\psi}_{i}|\dot{\psi}_{i}\rangle 
 = 
\frac{1}{2}\sum_{I,\alpha}\Delta M_{I}
                  \langle (v_I^{\alpha})^{2}\rangle
\label{eqn:temp_e}
\end{equation}
In other words if the electronic orbitals move rigidly with the ions
the actual inertia of the ions in a CP simulation can be obtained 
by adding to the ``bare'' ionic inertia the inertia carried by the
electronic orbitals.  
This result
has been pointed out previously \cite{blochl2} and ionic masses are commonly 
renormalized when dynamical
quantities are being investigated.
We now explore the consequences that such a modification of the ionic 
inertia has on typical observables extracted from CP simulations.
First, as already mentioned, the correct definition 
of temperature in a microcanonical CP simulation is given by
(\ref{eqn:realtemp}). Similarly, in a simulation 
where temperature is controlled, e.g. through a Nos\'e thermostat\cite{nose}, 
the quantity to be monitored corresponds to the 
instantaneous value of (\ref{eqn:realtemp}). 
Dynamical observables will also be affected by the additional 
inertia, as already noted in the case of phonons extracted from 
CP-MD in carbon systems \cite{kohanoff,scandolo}. 
In the case of
homogeneous systems (a single atomic species in which all the atoms 
are in similar local chemical environments) all dynamical 
quantities can be simply rescaled using the mass correction of eqn. 
(\ref{eqn:masscorr}). However, for heterogeneous systems the correction is not
always trivial, as different mass corrections apply to different atomic 
species due to different atomic kinetic energies.
In practice, we found that a convenient and more general way to express 
the mass correction of ion $I$ is given by 
\begin{equation}
\Delta M_{I} = f_{I}\frac{2m_{e}E_{k}^{total}}{3N\hbar^{2}}  
\label{eqn:deltaM2}
\end{equation}
where $f_{I}$ is a dimensionless constant which takes into account
the relative contribution of species $I$ to the total quantum
kinetic energy $E_{k}^{total}$.
 
\section{Simulations}
In order to gain more insight into the theory put forward in the previous 
sections, we have performed CP simulations on pressurised magnesium oxide 
and on silicon. Among the insulators (we restrict our analysis to 
insulators as adiabatic decoupling is less obvious in metallic systems and
this would complicate considerably our analysis), MgO and Silicon 
are extremal cases:
MgO is a highly ionic system with large quantum kinetic energy
associated with the strongly localised charge distribution; Silicon 
on the other hand is a covalent system where electron states are much more 
delocalised. Within our pseudopotential description
of MgO \cite{TM}, 
the 1s,2s and 2p states are frozen into the core of Mg whereas only the 1s states
are frozen into the core of O. Since there is very nearly complete transfer of the two 3s electrons
from Mg to O (inspection of charge density contour plots reveal no evidence of any valence charge 
anywhere except surrounding O sites) the electron quantum kinetic energy
may to a first approximation be attributed to electronic states localised on oxygen ions.
This makes MgO an ideal system to study within the rigid ion model since only the oxygen mass
will be rescaled. As mentioned in the previous section, additional 
problems arise if one deals with more than one electron-carrying 
species as the quantum kinetic energy must be divided between these species.
The large quantum kinetic energy of MgO means that the error in the CP forces should
be large relative to many materials. The simulations of MgO were performed at a high
pressure ($\sim 900$ kbar) as this enhanced its ionicity.

Silicon on the other hand is a covalent/metallic  system with relatively low quantum kinetic
energy. As such it should be one of the systems most favourable to the Car-Parrinello 
approximation but least favourable to description in terms of rigid ions.

\subsection{Technical Details}
We have performed ten different simulations. The technical details 
are summarised in table ~\ref{table:details}. 

All simulations were performed with a cubic simulation cell  
of side $L$ (see table~\ref{table:details}) under periodic boundary conditions and with 64 atoms 
in the unit cell. We used a plane wave basis set with an energy cut off for the wavefunctions of
$E_{cut}$. The Brillouin zone was sampled using only the $\Gamma-$point.
In each simulation we have used the mass preconditioning scheme of Tassone {\em et al.}\cite{tassone}
and the parameters $\mu_{0}$ and $E_{p}$ in table~\ref{table:details} are defined as in Ref.[5].

Liquid silicon is metallic and so, as suggested
by Bl\"{o}chl and Parrinello ~\cite{blochl}, two Nos\'{e} thermostats were used to counteract
the effects of energy
transfer between the ions and the $|\delta\psi_{i}\rangle$ due to overlap of their
frequency spectra.The values of the parameters used were
$Q_{e} = 21.3$ a.u./atom,$E_{kin,0}=1.65\times 10^{-4}$ a.u./atom and $Q_{R} =244400$ a.u..
These were chosen for compatibility with those of Ref.[10] by taking
into account the slight increase in temperature and scaling accordingly.

In all simulations, with the exception of simulation 2, the system
was first allowed to evolve for at least $1$ ps and this trajectory was discarded. 
For simulation 2, this initial equilibration time was $0.5$ ps.
All results reported are taken from the continuations of these equilibration
trajectories.

In all simulations, the total quantum kinetic energy of the system, and hence the
average mass correction (\ref{eqn:deltaM2}), varied during the simulation by less than $0.3\%$. It was therefore
taken as a constant in further analysis.

The total energy of all the degrees of freedom (including the thermostats 
in simulation 3) 
was conserved in all simulations {\em at least} to within one part in $10^{5}$.

\subsection{Results}
In order to check the predictions of the theory developed in 
Sections II and III, we have taken segments of CP trajectories 
and calculated the true BO forces along these segments by 
putting the electronic orbitals to their ground state with 
a steepest descent method.
We look at the instantaneous error in the $\alpha^{\scriptstyle th}$ 
cartesian component of the CP force on atom $I$ relative to the 
root-mean-squared (r.m.s) BO force component, i.e :

\begin{eqnarray}
\delta F_{I}^{\alpha}(t)
& = &
\frac{
\Delta F_{I}^{\alpha}(t) }{\sqrt{
\frac{\sum_{c}\sum_{J,\beta} (F_{BO_{J}}^{\beta})^{2} }
{3 N N_{c}}}}
\label{eqn:df1}
\end{eqnarray}
and the instantaneous relative error minus the relative error 
predicted by the rigid-ion model :
\begin{eqnarray}
[\delta F_{I}^{\alpha}(t)]_{corr}
 =
\frac{
\Delta F_{I}^{\alpha}(t)+\Delta M_{I} \ddot{R}_{I}^{\alpha} }{\sqrt{
\frac{\sum_{c}\sum_{J,\beta} (F_{BO_{J}}^{\beta})^{2} }
{3 N N_{c}}}
}
\label{eqn:df2}
\end{eqnarray}
where $N_{c}$ is the number of ionic configurations at which the error
in the CP forces was calculated and $\sum_{c}$ is the sum over all such 
configurations. The value of $\Delta M_{I}$ in (\ref{eqn:df2}) is determined 
using the rigid-ion-model expression (\ref{eqn:deltaM2}), and $E_{k}^{total}$
was given its average value during the simulation. 
The scaling parameters which were found to give best results for silicon and oxygen
were $f_{Si}=1.0$ and $f_{O} = 1.92$ respectively.

We also look at $\langle \delta F_{I}^{\alpha}\rangle_{ } $ and
$\langle \delta F_{I}^{\alpha}\rangle_{corr} $ the r.m.s
values of $[\delta F_{I}^{\alpha}]_{ }$ and $[\delta F_{I}^{\alpha}]_{corr}$
over all the ions, cartesian components and
configurations tested.

Since the CP forces are affected by a ``fast'' component whose effect 
on the ionic dynamics is believed to average out on the time scale of the
ionic motion, we introduce the quantity $\Gamma(\tau)$, defined as
\begin{equation}
\Gamma(\tau) = \int_{0}^{\tau}\frac{1}{3N}\sum_{I,\alpha}\Bigg|\frac{
\int_{t_{0}}^{t_{0}+\tau}\Delta F_{I}^{\alpha}(t)dt}
{\int_{t_{0}}^{t_{0}+\tau}|\Delta F_{I}^{\alpha}(t)|dt}\Bigg | d\tau \label{eqn:gamma}
\end{equation}
If we begin our comparison between CP and BO forces at some instant $t_{0}$
along the trajectory then inspection of $\Gamma(\tau)$ gives a feeling for
how large the fast component is.
If the errors in all the forces of the system oscillate rapidly with an average of zero then
$\Gamma(\tau)$ decreases very quickly from the value of one at $\tau=0$ to
zero at $\tau\sim \tau_e$. 
For systematic errors $\Gamma(\tau)$ should decrease gradually from one to zero on a timescale
of the order of the period of $\tau_i$. In realistic cases $\Gamma(\tau)$ drops from 
one and levels off to a smaller value for $\tau \sim \tau_e$, and then decreases gradually to zero
for $\tau$ exceeding $\tau_i$. 
The value of $\Gamma(\tau)$ on the plateau between $\tau_e$ and $\tau_i$
provides a measure of how 
much the errors calculated in (\ref{eqn:df1}) and (\ref{eqn:df2}) are 
attributable to a systematic (i.e. ``slow'') departure from the BO surface.

We begin by looking at the forces in Silicon in both the solid at $ 330$ K and
the liquid at $ 2000$ K (simulations 1,2 and 3). Simulation 1 was preceded
by a short run where the temperature was set to about 1000 K. Electrons were
then relaxed in their ground state and the ionic velocities
set to zero. This allows the electrons to smoothly accelerate with the ions. 
A microcanonical simulation followed where the ionic temperature reached,
after a short equilibration, the value of 330 K. This procedure was 
followed in all the simulations reported here, except where discussed. 

\begin{figure}[h]
\begin{center}
\includegraphics[width=8cm,height=12cm]{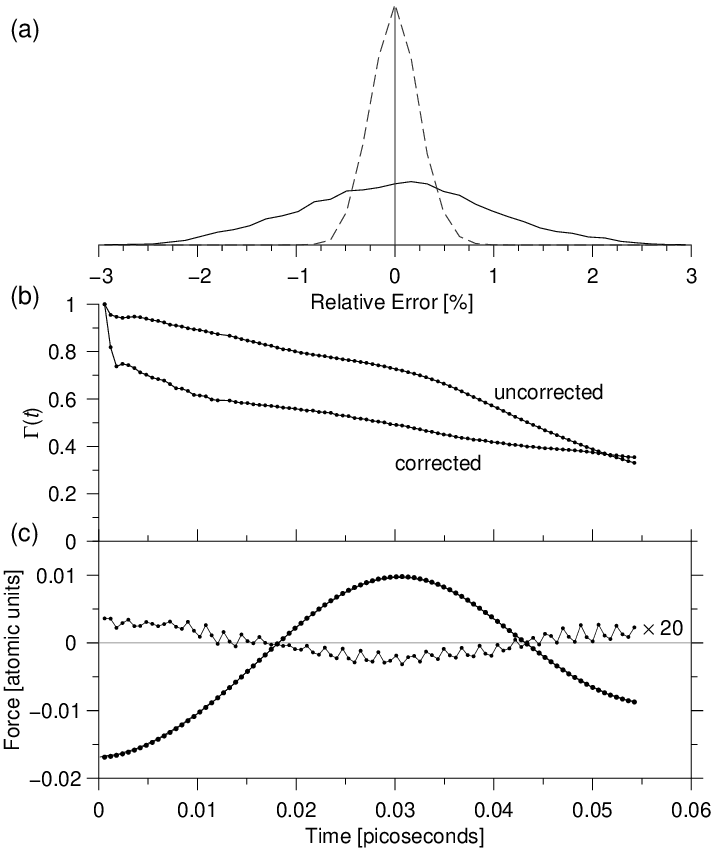}
\\
\caption{Simulation 1.
(a) Distribution among all atoms $I$ and all cartesian components $\alpha$ 
of the percentage errors in the CP forces relative
to the BO forces at the same ionic positions, $100\times\delta F_{I}^{\alpha}(t)$ (full line) 
and these errors when the forces have been partially corrected according to a rigid-ion model,
 $100\times[\delta F_{I}^{\alpha}(t)]_{corr}$ (dashed line). 
(b) $\Gamma(t)$ as defined by Eqn. ~\ref{eqn:gamma} for the full error in the forces and those
as partially corrected according to the rigid-ion model (c) 
$F_{BO_{I}}^{\alpha}$,$F_{CP_{I}}^{\alpha}$ and $(F_{CP_{I}}^{\alpha}-F_{BO_{I}}^{\alpha})$
(multiplied by a factor of $20$ for visibility)
for a typical force component.
Dots indicate the points at which
the BO force was calculated (every 5 time steps).
}
\label{fig:si300ksmooth}
\end{center}
\end{figure}

In solid Si at $330$ K (Fig. ~\ref{fig:si300ksmooth} )
we find that the standard deviation of the error in
forces is $0.94\%$. However, most of this error can be 
attributed to a rigid dragging of the Si atomic orbitals. 
The standard deviation of the error is in fact reduced to 
$0.24\%$ after the rigid-ion correction (\ref{eqn:masscorr}) 
is subtracted. The $\sim 30\%$ drop
of $\Gamma(t)$ (corrected) shown in Fig.~\ref{fig:si300ksmooth}b 
indicates that $\sim 30\%$ of the residual $0.24\%$
error can be attributed to ``fast'' oscillations, so that the overall 
average error introduced by the CP approximation, once corrected for the
rigid dragging and under the assumption that the fast component is not 
relevant, is less than $0.2\%$. 

As has been pointed out previously by Remler and Madden \cite{remler},
it is important to begin the dynamics with electrons and ions 
moving in a consistent way as we have done here in all simulations 
except the one we now discuss (simulation 2) and in the case of liquid Si 
(simulation 3).
We found that the error in the forces increases substantially 
if the simulation is
not started from zero ionic velocities, a procedure that would 
otherwise have the advantage of shortening considerably the time 
needed to reach thermal equilibrium. Simulation 2 started from the 
end of simulation 1, but electrons have been put in the ground state
before restarting (ionic velocities and positions were instead kept 
unchanged). Forces were tested after $0.5$ps from the electron quenching.

\begin{figure}[h]
\begin{center}
\includegraphics[width=8cm,height=12cm]{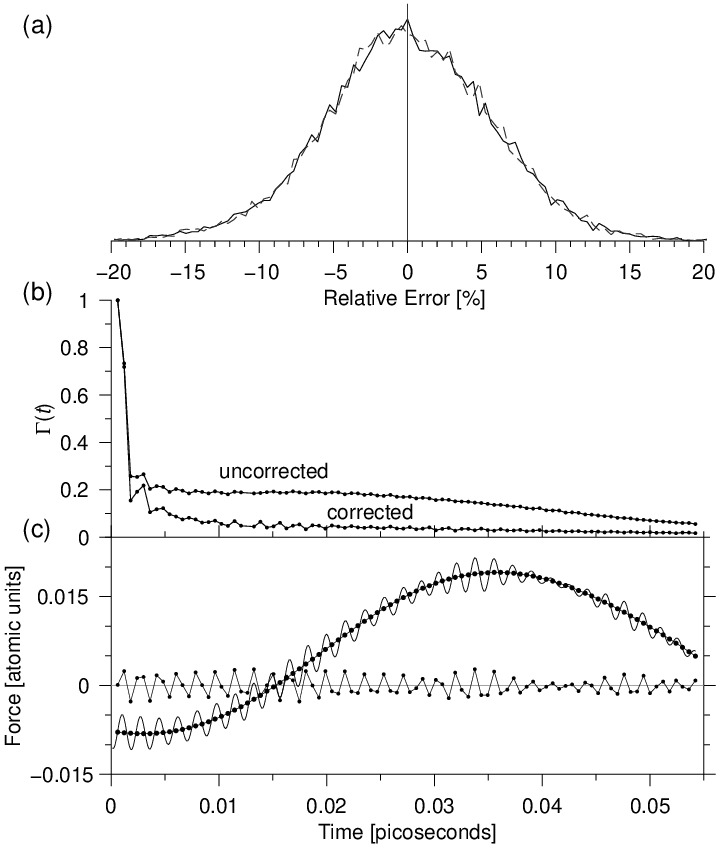}
\\
\caption{Simulation 2, Crystalline Si at $330$K when the electrons receive
a 'kick' at the beginning of the simulation. 
See caption of Fig. 1 for explanation. $(F_{CP_{I}}^{\alpha}-F_{BO_{I}}^{\alpha})$ 
has not been scaled for visibility.}
\label{fig:si300krough}
\end{center}
\end{figure}

The standard deviation of the error in forces is now $5.7\%$ and the
error in the forces as corrected according to the rigid ion approximation
at $5.68\%$, is not significantly improved. However clearly from 
inspection of $\Gamma(t)$ in Fig.~\ref{fig:si300krough}b and the
sample force component in Fig. ~\ref{fig:si300krough}c most of this 
error can be attributed to the high frequency oscillations of the 
electronic orbitals. If we assume that these oscillations do not 
influence the ionic dynamics, the error reduces to about 1.2\% 
for the uncorrected forces and to less than 0.5\% for the corrected
forces. The amplitudes of these oscillations are nevertheless 
significant and may affect the thermodynamics in a way that is not easy to
predict. These oscillations clearly originate from the initial jerk
experienced by the the electrons in their ground state and 
survive for a long time due to the adiabatic decoupling.

In the liquid the situation is considerably worse than in the crystal.
The standard deviation of the error is $3.4\%$ which improves only to $3.1\%$ with
the rigid-ion correction. There do not seem to be high frequency, 
high amplitude oscillations here despite the simulation being started 
with finite ionic velocities.
However, there are oscillations of a lower frequency (although 
still quite high relative
to ionic timescales) which are probably due to the presence of the 
Nos\'{e} thermostat.
It may be that the Nos\'{e} thermostat has the effect of damping 
out the kinds of
oscillations seen in Fig.~\ref{fig:si300krough} but the presence of these other
oscillations is hardly an improvement. 
This highlights the need for careful choice
of parameters for the Nos\'{e} thermostat, particularly the value of $E_{kin,0}$. It is
not clear how one should choose this parameter in general. For example, here we have used
a value of $E_{kin,0}$ compatible with Ref.[10], however we note that this
is considerably smaller than the value recommended in Ref.[6]
which was obtained according to a rigid ion model. We have also done simulations using
higher values of $E_{kin,0}$ and in all cases the errors in the forces have been greater.
It is likely therefore that by decreasing further $E_{kin,0}$ we might improve further the
forces however this has not been attempted here. 
The issue of thermostatting a system in a way which minimizes the kinds of errors seen here
while accounting for the evolution of the electronic ground state
in a more general way than is allowed by the rigid-ion approximation
has recently been tackled by Bl\"{o}chl\cite{blochl3}.

\begin{figure}[h]
\begin{center}
\includegraphics[width=8cm,height=12cm]{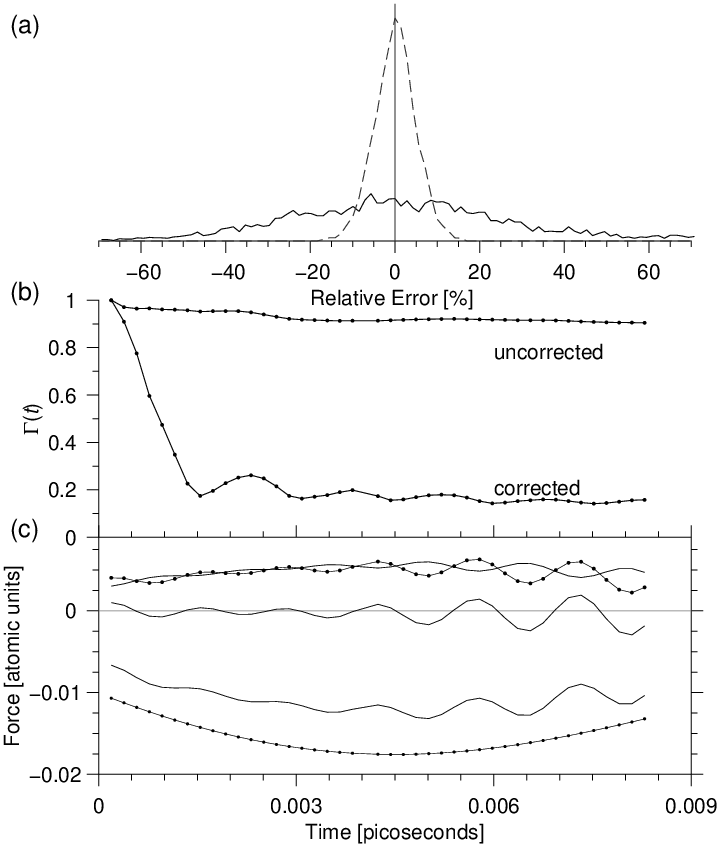}

\caption{Simulation 4. Forces on the oxygen ions in crystalline MgO at $2800$ K.
(a) and (b) are as in Fig. 1
(c) 
From top to bottom $(F_{CP_{I}}^{\alpha}-F_{BO_{I}}^{\alpha})$,$-\Delta M_{O}\ddot{R}_{I}^{\alpha}$(dotted
line),
$(F_{CP_{I}}^{\alpha}-F_{BO_{I}}^{\alpha}+
\Delta M_{O}\ddot{R}_{I}^{\alpha})$,$F_{CP_{I}}^{\alpha}$,$F_{BO_{I}}^{\alpha}$
for a typical force component. 
Dots indicate the points at which
the BO force was calculated (every time step).
}
\label{fig:O_solid400}
\end{center}
\end{figure}

We now look at the forces in crystalline MgO with $\mu_{0}=400$a.u (Fig.~\ref{fig:O_solid400}).
The relatively high quantum kinetic energy associated with states attached 
to the O ions means that, according to eq. (\ref{eqn:A}), 
the errors in the forces are considerably
larger for the O ions than we have seen for Si.
The errors in the CP forces have in fact a standard deviation as 
large as $32\%$ .  However, when this is corrected as in (\ref{eqn:masscorr})
by attributing all the quantum kinetic energy to states rigidly 
following the O ions
the standard deviation of the error reduces to $4.8\%$. 
Furthermore, the corrected value of $\Gamma(t)$ indicates that about $80\%$ of the error on the O forces
cancels out after account is taken for the high frequency oscillations, 
suggesting that a more appropriate estimate of the error is $\sim 0.4\%$.
The amplitude of the fast oscillations is a cause for concern however 
and since the simulation was begun at zero ionic velocity it is not clear 
how it may be reduced further.

\subsubsection{Temperature}
\begin{figure}[h]
\begin{center}
\includegraphics[width=10cm,height=8cm]{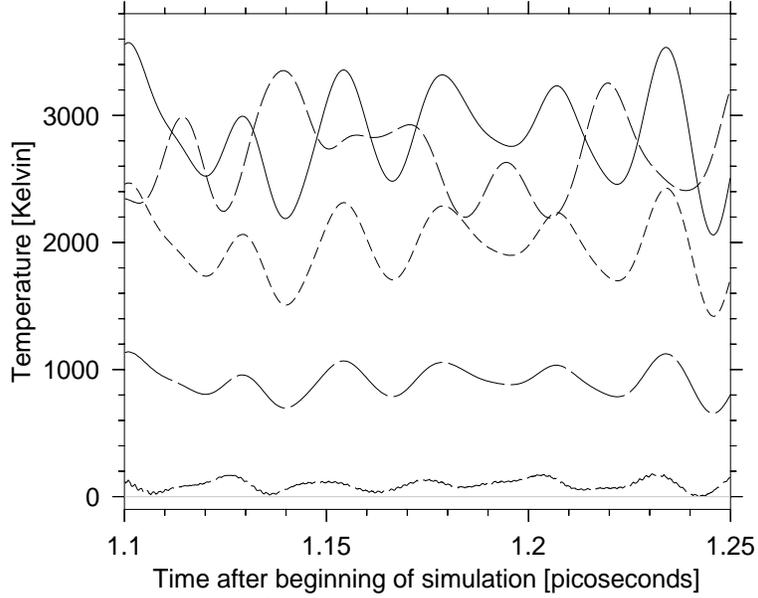}

\caption{Simulation 1. Dashed lines from top to bottom are :
 Mg temperature,
O temperature,
 $T_{el}$ ,
and $10\times(T_{el}-T_{\Delta M_{O}})$.
The full line is the Oxygen temperature when it is calculated with a mass
which is increased by $A_{O}$.
}
\label{fig:temperature}
\end{center}
\end{figure}

We focus here only on MgO, as the effects of the electronic dragging are 
enhanced. According to the results of Section III, we should expect
a difference between the naive definition of temperature and the 
definition corrected by the electronic dragging, eqn. (\ref{eqn:realtemp}). 
In the case of MgO,
as noted in the previous section, this correction affects only the
oxygen atoms, as only a minor amount of electronic charge is carried
by the Mg$^{2+}$ ion. In Fig. \ref{fig:temperature} we show the behavior
of the instantaneous values of the naive and corrected temperatures.
The corrected temperature exceeds the naive definition by about 500 K.
More interestingly, we report in Fig. \ref{fig:temperature} also the 
contributions to the temperature of the two atomic species. It is
clear that the naive definition would imply that the two species are not
at thermal equilibrium. On the other hand, use of the corrected definition
for the oxygen temperature brings the temperature of the two species
in much better agreement, supporting the conclusion, based on 
the rigid ion model, that thermodynamics can be restored 
by a simple rescaling of the oxygen mass. The mass rescaling, as calculated
with (\ref{eqn:error3}) amounts to $\Delta M_O \sim 7.5$ atomic mass units
($M_O = 16$ atomic mass units). 
We also report in Fig. \ref{fig:temperature} the instantaneous value 
of the fictitious electronic kinetic energy, the l.h.s. of eq. 
(\ref{eqn:temp_e}), and the difference between this quantity and the
r.h.s. of eq. (\ref{eqn:temp_e}), which represents the contribution due to
the rigid dragging of the electronic orbitals. The difference is very small,
implying that residual contributions due, for example, to the fast
electronic oscillations are negligible in MgO compared to the slow 
dragging of the orbitals.

\subsubsection{Phonon Spectra}
We have calculated the phonon densities of states of crystalline Si and MgO by fourier
transforming the velocity autocorrelation function.
In all cases, the first picosecond of the simulation was discarded and results 
obtained by averaging over at least one subsequent picosecond.
For Silicon the velocity autocorrelation function was calculated on a time domain
of length $1.2$ps and for MgO on a time domain of length $0.5$ps.

\begin{figure}[h]
\begin{center}
\includegraphics[width=10cm,height=8cm]{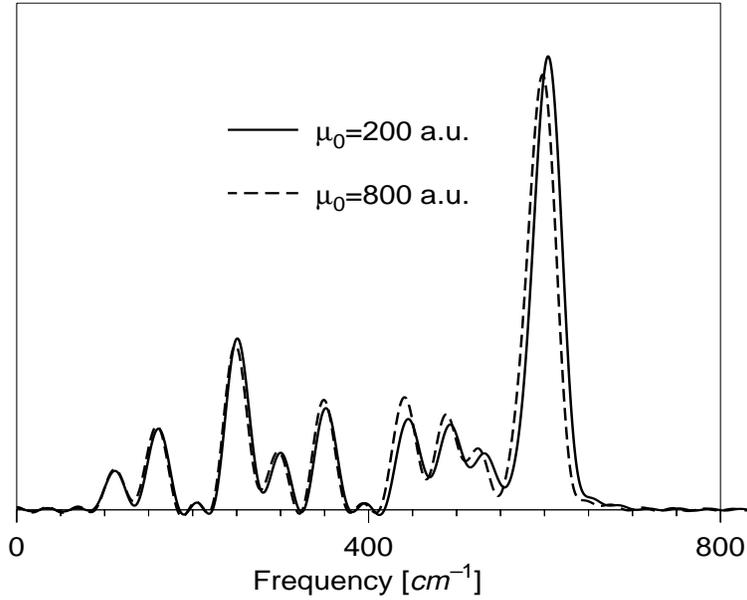}

\caption{Phonon density of states of crystalline Silicon for $\mu_{0} =200$ a.u. (simulation 5)
and for $\mu_{0} =800$ a.u (simulation 6). 
}
\label{fig:phonon_si}
\end{center}
\end{figure}

In silicon (Fig. ~\ref{fig:phonon_si}) the difference is reasonably small.
According to the rigid ion model the frequencies should be corrected using
\begin{equation}
\omega_{\tiny corrected} = \omega_{\tiny CP}\sqrt{ 1 + \Delta M/M}
\end{equation}
where $\omega_{\tiny CP}$ is the frequency as extracted directly from the 
CP simulation. 
We find that for silicon this overestimates by about a factor of two 
the amount of the correction. This small discrepancy may be due to the length
of simulation used for calculating the frequency spectra or due to a breakdown 
of the rigid-ion description when $\mu_{0} = 800 a.u.$. It may also be that
neglecting the effect 
of the fast oscillations is not be completely appropriate when the 
dragging contribution is small.   

\begin{figure}[h]
\begin{center}
\includegraphics[width=10cm,height=10cm]{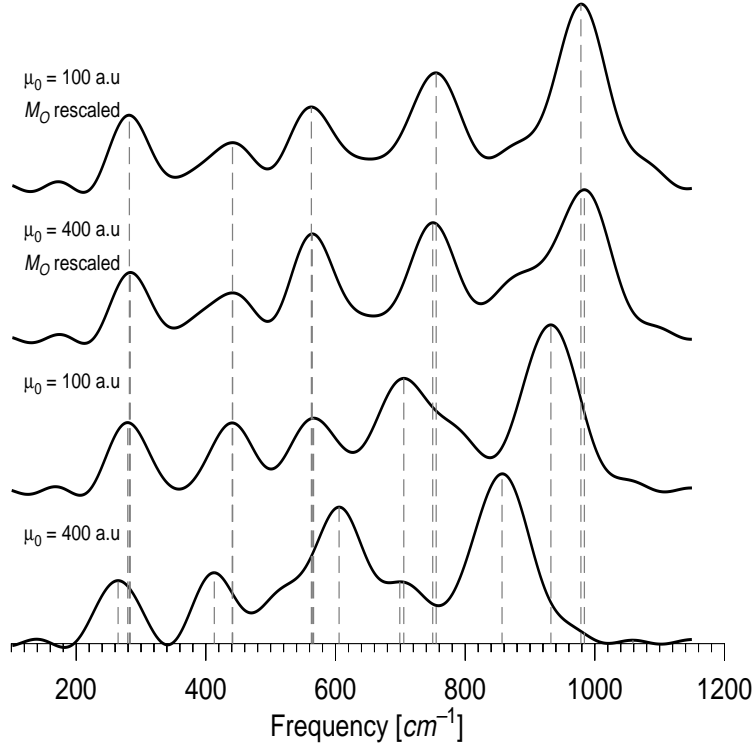}

\caption{Phonon density of states of crystalline MgO for $\mu_{0} =100$ a.u. 
and for $\mu_{0} =400$ a.u. with rescaled (simulations 7 and 8)  
and unrescaled (simulations 10 and 4) oxygen masses.
}
\label{fig:phonon_mgo}
\end{center}
\end{figure}

In MgO, as expected the difference is much larger. 
We calculate the phonon spectra for $\mu_{0}=400$a.u. and for 
$\mu_{0}=100$a.u and find large differences between them 
(see Fig. ~\ref{fig:phonon_mgo}) highlighting
again how the dynamics depends on the value of $\mu$.
The fact that two species
are involved complicates matters as the mass correction is different for the 
two species (it actually vanishes for Mg). Therefore we should not 
expect simply a rigid shift of the frequencies.
However, if the rigid-ion approximation is valid, one may conceive
to rescale the oxygen mass {\em a priori} in eq. (\ref{eqn:ions1})
as $\tilde{M}_{O} = M_{O} - \Delta M_{O}$, so that the 
actual CP dynamics expressed in terms of the BO forces, 
eq. (\ref{eqn:masscorr}), becomes identical to the BO dynamics
 if the rigid ion approximation holds.
We have done this for MgO, again for $\mu_{0}=400$a.u. and 
$\mu_{0}=100$a.u and we see that
the results are much improved. There are only small 
differences in the positions of the
peaks and the overall shapes of the curves are very similar.

\subsubsection{Dependence of error on $\mu$}
\begin{figure}[h]
\begin{center}
\includegraphics[width=8cm,height=6cm]{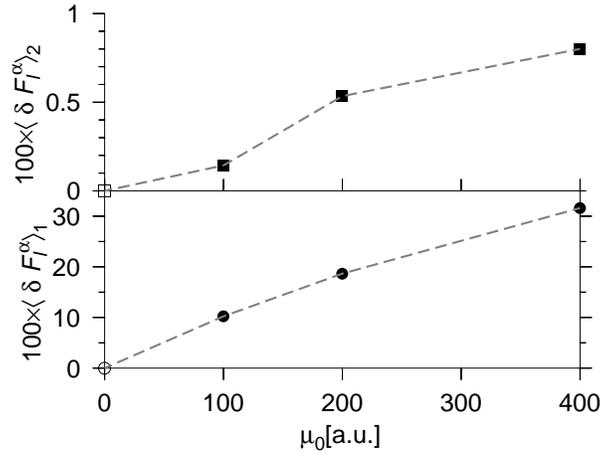}

\caption{Scaling of the standard deviation of the errors in the forces on the oxygen ions with $\mu_{0}$
(Simulations 4,9 and 10).
$\langle \delta F_{I}^{\alpha}\rangle_{corr} $ has been reduced to eliminate cancelling
high frequency oscillations by inspection of $\Gamma(t)$.
}
\label{fig:scaling}
\end{center}
\end{figure}

We now try to address the question of how the error in the 
CP forces depends on the fictitious electron mass. 
According to Eqn.~\ref{eqn:error2} the error should scale approximately 
linearly with the mass. However this is based on the simplifying assumptions
that the oscillations in $|\delta\psi_{i}\rangle$ have a small amplitude 
and that the $|\delta\psi_{i}\rangle$ do not on average exchange energy 
with the ions, i.e. full adiabatic decoupling is achieved. 
Fig. ~\ref{fig:scaling} shows $\langle \delta F_{I}^{\alpha}\rangle$
and $\langle \delta F_{I}^{\alpha}\rangle_{corr} $ for the oxygen ion for
three different values of $\mu_{0}$ where $\langle \delta F_{I}^{\alpha}\rangle_{corr} $
has been scaled to eliminate the contribution of errors from high frequency oscillations
by inspection of $\Gamma(t)$. 

Since the uncorrected error is dominated by the effect of the displacement
of the equilibrium positions of the orbitals from the ground state, this
scales approximately linearly with $\mu_{0}$. The small error which remains
after the rigid ion correction has been applied could have contributions 
from many different sources including deviations from the rigid ion 
description and interactions between the ions and the 
$|\delta \psi_{i}\rangle$.  It is also of the order of the fluctuations in $E_{k}^{total}$
during the simulation.

\section{Discussion}
The various cases studied in this work have been explicitly chosen 
because of their ``extremal'' behavior. 
The oxygen ions in MgO have a very strong electron-ion coupling as 
shown by the large quantum kinetic energy.  However, oxygen seems to be 
well described within the rigid-ion approximation and so the 
thermodynamics are probably quite close to those of a BO system. 
Crystalline silicon is less well described in terms of rigid ions 
(although still remarkably well) but it has a much lower electron-ion coupling
so the errors in the forces are very small. If we ``measure'' the departure
of the CP dynamics from the BO dynamics in terms of $\Delta M / M$, with
$\Delta M$ defined as in (\ref{eqn:masscorr}), then it appears that the elements
where the departure is expected to be larger are located in the upper right
of the periodic table, because they combine a low atomic mass with a large
binding energy of the valence electrons (and thus a large quantum kinetic
energy). Transition metals may also be strongly affected, because of the large
number and strong localisation of the $d$-electrons. However, the higher 
the localisation of the orbitals, the higher the chances that the description of
the electronic dynamics in terms of rigid orbitals is correct. The 
large departure observed in the case of MgO suggests that a proper assessment 
of how much the CP forces differ from the BO ones is mandatory in most 
systems. This can be achieved by either calculating the BO forces for 
selected ionic configurations, or by performing simulations for 
different (smaller) values of $\mu$, and checking how the results 
scale with decreasing $\mu$. If the departure is large then it is likely 
that in many cases the CP forces can be brought into good agreement with the BO ones
by simply rescaling the ionic masses. 

Additional complications may arise when the dynamics lead to fundamental changes
in the electronic structure. The first and second derivatives of the electronic orbitals
with respect to the positions of the ions, which appear in equation ~\ref{eqn:error2}, 
may become relevant in regions of phase space where
the electrons play a significant role. For example, 
if charge transfer between ions occurs, or if a substantial 
rearrangement of the electronic orbitals takes place, as in a chemical 
reaction,
then the simple method of rescaling the ionic masses will no longer work.

If one is to judge the quality of a CP simulation by the errors in the forces
as we have largely done here, then
a question which needs to be addressed is to what extent these errors manifest
themselves as errors in the properties of interest in the simulation.
It is likely that random high frequency oscillations in the forces such as those due to 
the dynamics of the $\delta\psi_{i}$ have no discernible effect on the thermodynamics
of the system if such oscillations are small. 
The magnitude of the oscillations seen here in the case of 
MgO is a cause for concern however. 

A further problem may arise at high temperatures where, if there is 
electron-ion coupling, energy can pass between ions and electronic orbitals
and the ionic and orbital subsystems may begin to equilibrate 
leading to an increase
in the fictitious electronic kinetic energy.
This can be seen by inspection of equation ~\ref{eqn:electrons4}, where the ionic terms
may vary over short timescales if the ionic kinetic energy is very high
and the fictitious mass $\mu$ too large.
It has previously been supposed\cite{buda,blochl} that such an 
irreversible increase of kinetic 
energy was due to overlap of the ionic frequency spectrum with that of the electrons
due to a small or vanishing bandgap, $E_{g}$. However a small bandgap is a sufficient but 
not a necessary condition
for this energy transfer to occur.

All of the effects discussed in this paper are dependent on the choice of the fictitious
mass parameter, $\mu$, and by reducing this parameter all thermodynamic and dynamic
properties of a simulation may be brought arbitrarily close to those in a Born-Oppenheimer
system. However, a reduction of $\mu$ has the drawback that the time step required
to integrate the equation of motion for the electronic orbitals is reduced thereby
decreasing the computational efficiency of the method.
By checking how the property of interest in a simulation scales with $\mu$ one 
can control the level of approximation with which it is calculated.

\section{Conclusions}
Under the assumption that high frequency electronic oscillations
(i.e. the dynamics of the $|\delta\psi_{i}\rangle$)
are small and independent of ionic motion, we have shown that 
Car-Parrinello simulations amount to 
solving the equation of motion for the ions 
\begin{eqnarray}
 M_{I}\ddot{R}_{I}^{\alpha} = F_{BO_{I}}^{\alpha} +
2\sum_{i}\mu_{i}\Re\bigg \{\sum_{J}\ddot{R}_{J}^{\beta}\frac{\partial \langle \psi_{i}^{(0)}|}{\partial R_{I}^{\alpha}}
\frac{\partial|\psi_{i}^{(0)}\rangle}{\partial R_{J}^{\beta}}
+ \sum_{J,K} \dot{R}_{J}^{\beta}\dot{R}_{K}^{\gamma}\frac{\partial \langle \psi_{i}^{(0)}|}{\partial R_{I}^{\alpha}}
\frac{\partial^{2}|\psi_{i}^{(0)}\rangle}{\partial R_{K}^{\gamma}\partial R_{J}^{\beta}}\bigg\}
\end{eqnarray}

We have compared the forces in simulations of Si and MgO for a number of 
different values of $\mu$ to the BO forces and found that
in the case of Si the errors are small and change very slightly the
phonon spectrum of the crystal. 
In MgO we observe very large systematic errors in the forces which are however 
mostly
attributable to a rescaling of the mass of the oxygen ion thereby preserving 
the thermodynamics. When corrected for this effect the errors are slightly higher than 
those in crystalline Si but still quite small. The phonon densities of states further
confirm both the inadequacy of CP at these values of $\mu$ without the rigid-ion correction 
to describe dynamics and the ability of the rigid-ion model to correct the dynamics
in MgO.

We have demonstrated the necessity for checking the dependence of results of CP simulations
on the value of the fictitious mass parameter $\mu$.

{\bf Acknowledgements.} 
The authors would like to thank R. Car and N. Marzari for useful discussions.
We also thank P. Bl\"{o}chl for providing us with an advance copy of his manuscript\cite{blochl3}.
In addition P.T. would like to thank A. Trombettoni for useful discussions and M. Payne
for some valuable suggestions.
In this work extensive use was made of the Keck Computational Materials Science Laboratory
at the Princeton Materials Institute.

 \begin{table}
   \caption{Technical Details of the Simulations}
    \begin{tabular}{|c|c|c|c|c|c|c|c|c|}
Simulation \# & System & Temperature & $\mu_{0}$
& $E_{p}$ & $E_{cut}$ & $\Delta t$ & $\sum_{i}\mu_{i}\langle\dot{\psi}_{i}|
\dot{\psi}_{i}\rangle$& L\\\cline{3-9}
& & Kelvin & a.u. & Ryd. & Ryd. & a.u. & a.u.$\times 10^{4}$& a.u.
\\
\hline
 1 & Si & 330 & 270 & 1.0 & 12.0 & 5.0 &4.36&20.42\\
 2 & Si & 330 & 270 & 1.0 & 12.0 & 5.0 &4.36&20.42\\
 3 & Si(liquid) & 2000 & 270 & 1.0 & 12.0 &10.0 &4.35&19.8\\
 4 & MgO & 2800 & 400 & 2.7 & 90.0& 8.0 & 66.3&14.5\\
 5 & Si & 330 & 200 & 1.0 &12.0 & 5.0&3.23&20.42\\
 6 & Si & 330 & 800 & 1.0 &12.0 & 10.0&12.92&20.42\\
 7 & MgO ($M_{O}$ rescaled)& 2800 & 100 & 2.7 &90.0& 4.0 &16.55&14.5\\
 8 & MgO ($M_{O}$ rescaled)& 2800 & 400 & 2.7& 90.0& 8.0&66.4&14.5\\
 9 & MgO & 2800 & 200 & 2.7  & 90.0& 5.65& 33.1&14.5\\
 10& MgO & 2800 & 100 & 2.7  & 90.0& 4.0 &16.55&14.5\\
    \end{tabular}
\label{table:details}
 \end{table}

\begin{figure}[h]
\begin{center}
\includegraphics[width=8cm,height=12cm]{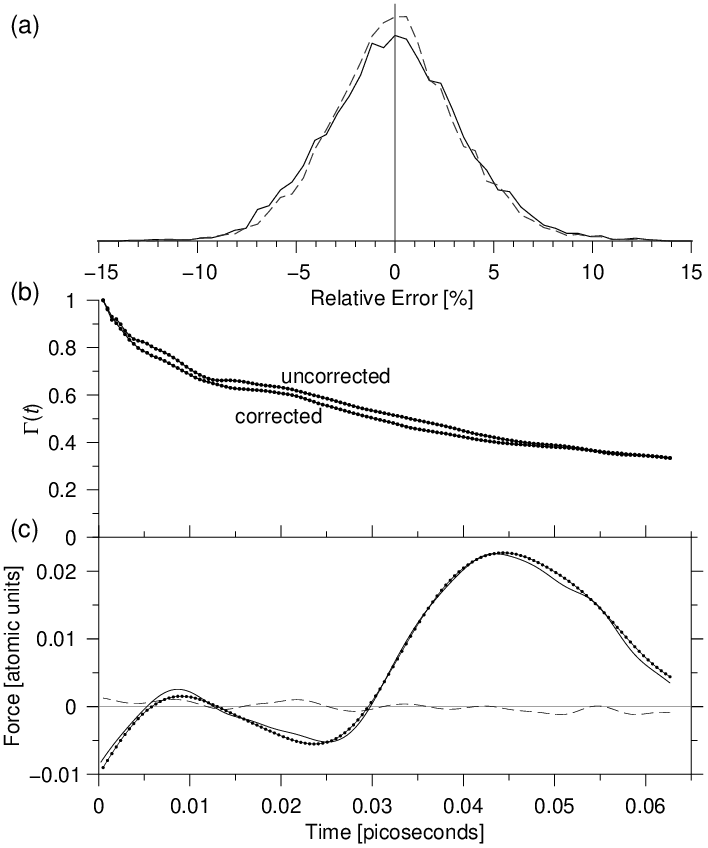}

\caption{Simulation 3, Liquid Si at $2000$ K. 
See caption of Fig. 1 for explanation. $(F_{CP_{I}}^{\alpha}-F_{BO_{I}}^{\alpha})$ 
has not been scaled for visibility.}
\label{fig:siliquid}
\end{center}
\end{figure}

\end{document}